\PassOptionsToClass{10pt}{revtex4-1}
\documentclass[aps,prl,showpacs,floatfix,twocolumn,byrevtex,superscriptaddress]{revtex4-1}
\usepackage{amsmath}
\usepackage{amssymb}
\usepackage{amstext}
\usepackage{amsopn}
\usepackage{amsfonts}
\usepackage{amsxtra}
\usepackage[english]{babel}
\usepackage{graphicx}
\usepackage{bm}
\usepackage{multirow}
\usepackage{dcolumn}
\usepackage{color}
\usepackage{hyperref}
\begin{document}

\title{\boldmath Orbital-differentiated coherence-incoherence crossover identified by photoemission spectroscopy in \boldmath LiFeAs \unboldmath}
\author{H. Miao}\email[]{Contributed equally to this work}
\affiliation{Beijing National Laboratory for Condensed Matter Physics, Institute of Physics, Chinese Academy of Sciences, Beijing 100190, China}
\affiliation{Condensed Matter Physics and Materials Science Department, Brookhaven National Laboratory, Upton, New York 11973, USA}
\author{Z. P. Yin}\email[]{Contributed equally to this work}
\affiliation{Department of Physics and Astronomy, Rutgers University, Piscataway, New Jersey 08854, USA}
\affiliation{Department of Physics and the Center of Advanced Quantum Studies, Beijing Normal University, Beijing 100875, China}
\author{S. F. Wu}
\affiliation{Beijing National Laboratory for Condensed Matter Physics, Institute of Physics, Chinese Academy of Sciences, Beijing 100190, China}
\author{J. M. Li}
\affiliation{Beijing National Laboratory for Condensed Matter Physics, Institute of Physics, Chinese Academy of Sciences, Beijing 100190, China}
\author{J. Ma}
\affiliation{Beijing National Laboratory for Condensed Matter Physics, Institute of Physics, Chinese Academy of Sciences, Beijing 100190, China}
\author{B. -Q. Lv}
\affiliation{Beijing National Laboratory for Condensed Matter Physics, Institute of Physics, Chinese Academy of Sciences, Beijing 100190, China}
\author{X. P. Wang}
\affiliation{Beijing National Laboratory for Condensed Matter Physics, Institute of Physics, Chinese Academy of Sciences, Beijing 100190, China}
\affiliation{Collaborative Innovation Center of Quantum Matter, Beijing 100190, China}
\affiliation{Department of Physics, Tsinghua University, 100084, Beijing, China}
\author{T. Qian}
\affiliation{Beijing National Laboratory for Condensed Matter Physics, Institute of Physics, Chinese Academy of Sciences, Beijing 100190, China}
\author{P. Richard}
\email[]{p.richard@aphy.iphy.ac.cn}
\affiliation{Beijing National Laboratory for Condensed Matter Physics, Institute of Physics, Chinese Academy of Sciences, Beijing 100190, China}
\affiliation{Collaborative Innovation Center of Quantum Matter, Beijing 100190, China}
\author{L. -Y. Xing}
\author{X. -C. Wang}
\affiliation{Beijing National Laboratory for Condensed Matter Physics, Institute of Physics, Chinese Academy of Sciences, Beijing 100190, China}
\author{C. Q. Jin}
\affiliation{Beijing National Laboratory for Condensed Matter Physics, Institute of Physics, Chinese Academy of Sciences, Beijing 100190, China}
\affiliation{Collaborative Innovation Center of Quantum Matter, Beijing 100190, China}
\author{K. Haule}
\affiliation{Department of Physics and Astronomy, Rutgers University, Piscataway, New Jersey 08854, USA}
\author{G. Kotliar}
\affiliation{Department of Physics and Astronomy, Rutgers University, Piscataway, New Jersey 08854, USA}
\affiliation{Condensed Matter Physics and Materials Science Department, Brookhaven National Laboratory, Upton, New York 11973, USA}
\author{H. Ding}
\email[]{dingh@iphy.ac.cn}
\affiliation{Beijing National Laboratory for Condensed Matter Physics, Institute of Physics, Chinese Academy of Sciences, Beijing 100190, China}
\affiliation{Collaborative Innovation Center of Quantum Matter, Beijing 100190, China}

\date{\today}

%
%

\begin{abstract}
In the iron-based superconductors (FeSCs), orbital differentiation is an important phenomenon, whereby correlations stronger on the $d_{xy}$ orbital than on the $d_{xz}$/$d_{yz}$ orbital yield quasi-particles with $d_{xy}$ orbital character having larger mass renormalization and abnormal temperature evolution. However, the physical origin of this orbital differentiation is debated between the Hund's coupling induced unbinding of spin and orbital degrees of freedom and the Hubbard interaction instigated orbital selective Mott transition. Here we use angle-resolved photoemission spectroscopy to identify an orbital-dependent correlation-induced quasi-particle (QP) anomaly in LiFeAs. The excellent agreement between our photoemission measurements and first-principles many-body theory calculations shows that the orbital-differentiated QP lifetime anomalies in LiFeAs are controlled by the Hund's coupling. 
\end{abstract}



\maketitle


Understanding the origin of the electronic correlations in high-temperature superconductors is a key step towards uncovering the pairing mechanism of their unconventional superconductivity\cite{Johnson2001,Shen2003,Valla1999,Richard2009,Norman1997,Norman1998}. Unlike the copper oxide superconductors, where the electronic correlations are controlled by the onsite Hubbard-$U$ interaction, the multi-orbital and multi-band nature of the FeSCs poses a strong challenge to forming a clear picture of their strong orbital-dependent electronic correlations. In one scenario the electronic correlations arise because the FeSCs are in close proximity to a Mott state or an orbital-selective Mott state\cite{Yi2013,Yi2015,Yu2011,Medici2014}, where the Fe 3d orbitals are decoupled from each other and therefore their correlation strength is controlled by the Hubbard-$U$ interaction as in the copper oxides\cite{Medici2014}. In an alternative scenario the electronic correlations come mainly from the formation of large fluctuating local moments due to the Hund's rule coupling\cite{Yin2011a,Yin2011b}. For this reason, the FeSCs are dubbed Hund's metals\cite{Yin2011a}. In this paper, we use high-resolution angle-resolved photoemission spectroscopy to measure the quasi-particle (QP) self-energy $\Sigma(k,\omega)$ and its temperature evolution, and compare them with first-principles density functional theory plus dynamical mean field theory (DFT+DMFT) calculations to clarify the origin of the many-body correlations in FeSCs.

High-quality single-crystals of LiFeAs were synthesized by the self-flux method\cite{Miao2014}. High-resolution ARPES data were recorded at the Institute of Physics, Chinese Academy of Sciences, using the He I$_{\alpha}$ (h$\nu$ = 21.218 eV) resonance line of a helium discharge lamp. The angular and momentum resolutions were set to 0.2$^{\circ}$ and 3 meV, respectively. ARPES polarization-dependent measurements were performed at the Dreamline of Shanghai Synchrotron Light Source using a Scienta D80 analyzer with energy and momentum resolutions set to 0.2$^{\circ}$ and 10 meV, respectively. The photon energy 74 eV is selected to tune the $k_{z}$ = 0. To select specific orbitals, we employed linearly polarized light to the mirror plane of the sample. All samples were cleaved in situ. The data were recorded in a vacuum better than $3\times 10^{-11}$ Torr with a discharge lamp and 10$^{-10}$ Torr with synchrotron light source. Our calculations are performed using an ab initio theoretical method for correlated electron materials, based on a combination of DMFT and density functional theory (DFT)\cite{Yin2011b,Haule2010}. The calculation is performed at $K_{z}$ close to the experimental value. This computational method improves the DFT description of the electronic structure of FeSCs, predicts the correct magnitude of the ordered magnetic moments\cite{Yin2011a}, and improves the description of electronic spectral functions, Fermi surfaces\cite{Yin2011a,Yin2011b}, charge response functions such as the optical conductivity\cite{Yin2011b}, and spin dynamics\cite{Yin2014}. For sake of consistency, we use in this work the same crystal structure, Hubbard $\mathbf{U}$=5.0 eV, Hund's $J_{H}$=0.8 eV as in previous work\cite{Yin2014,Yin2011a}.
%
\begin{figure}[tb]
\includegraphics[width=\columnwidth]{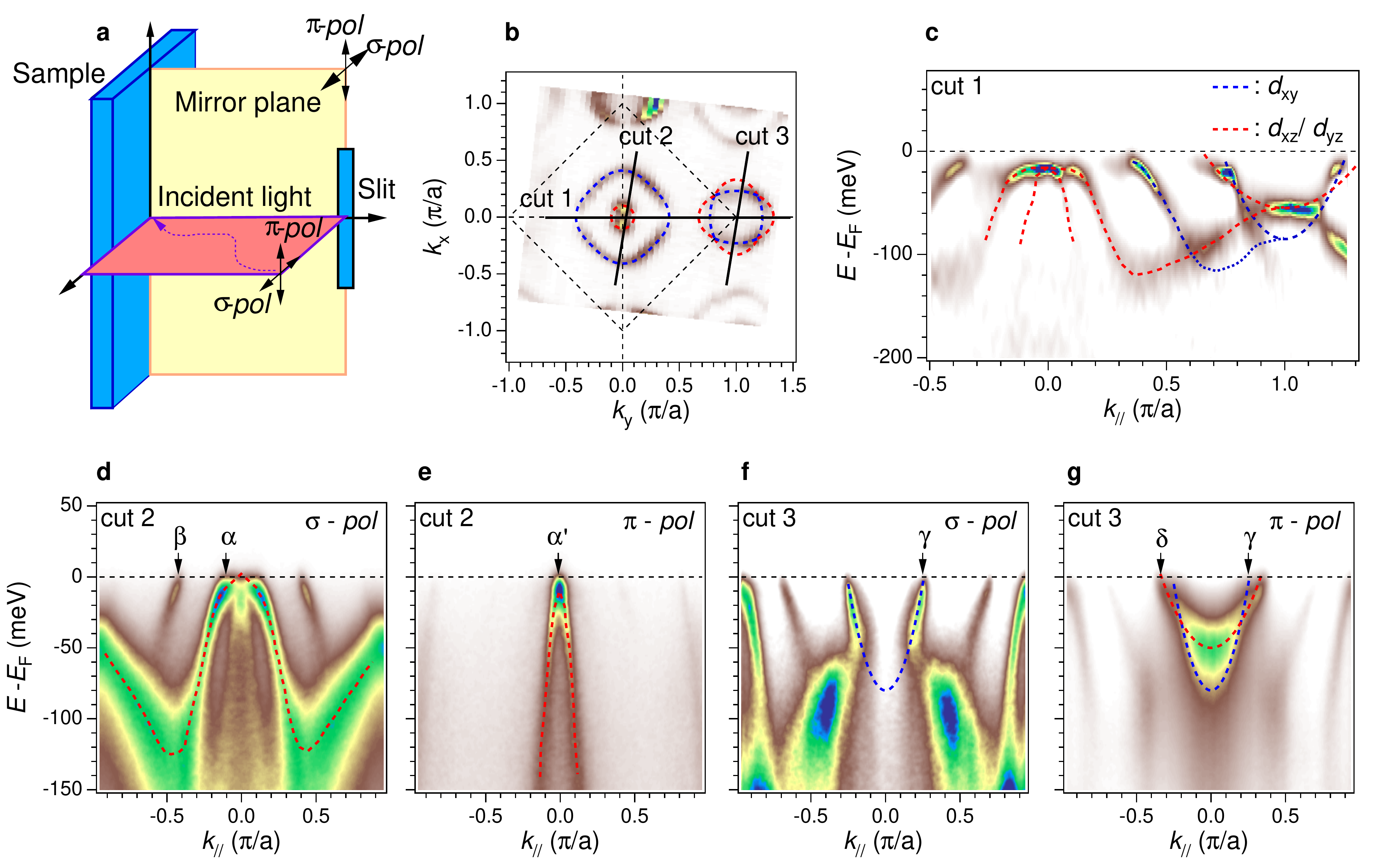}
\caption{ (color online) (a) Experimental setup of the polarization measurements. The $\pi$-polarization ($\pi$-pol) is pure, whereas the $\sigma$-polarization ($\sigma$-pol) is mixed with c-axis polarization. (b) FS mapping at 74 eV with the $\sigma$-pol. Dashed lines are the extracted FSs. The blue and red colors represent $d_{xy}$ and $d_{xz/yz}$ orbitals, respectively. (c) Curvature of ARPES intensity of cut1. The red and blue dashed dispersions are extracted from the high-resolution polarization data shown in (d)-(g).}
\label{Fig1}
\end{figure}

LiFeAs ($T_{c}$ = 18 K) offers an excellent platform for the study of QPs and the underlying many-body correlations. First, it is free of doping impurities, disorder, and has well-separated band dispersions\cite{Miao2014,Miao2015,Ye2014}, which is crucial to directly study the intrinsic QP dynamics. Second, unlike iron chalcogenides, LiFeAs has a single phase and has no complication of magnetic or orbital long-range orders at low temperature\cite{Dai2015}. Last but not least, first-principles calculations find that although the strength of the Hund’s coupling $J_{H}$ is similarly strong in iron-pnictides and iron-chalcogenides\cite{Yin2011a}, the band renormalization factor on the hole-like $d_{xy}$ band is 2-3 times smaller in the iron-pnictides than in the iron-chalcogenides. The similarities and differences between the pnictides and the chacolgenides are important for identifying the key correlations in all FeSCs, which are vital to understand the emergence of superconductivity and its pairing symmetry.

The electronic structure of LiFeAs consists of five bands near the Fermi energy ($E_{F}$) with three hole bands at the Brillouin zone (BZ) center and two electron bands at the corner of the 2 Fe/unit cell BZ. In order to determine their main orbital characters, we employ linearly polarized light to the mirror plane of the samples, as shown in Fig. 1a. Under this geometry, the $\pi$-polarization ($\pi$-pol) is pure and selects orbitals that have an even symmetry with respect to the mirror plane, while the $\sigma$-polarization ($\sigma$-pol) is mixed with the c-axis polarization, hence selecting orbitals that have an odd symmetry as well as the $d_{z^{2}}$ and $p_{z}$ orbitals\cite{Wang2012}. Fig. 1b shows the Fermi surface (FS) mapping at 74 eV with the $\sigma$-pol, the red and blue dashed lines corresponding to the extracted FSs. The curvature of the ARPES intensity of cut1 is displayed in Fig. 1c. The red and blue dashed dispersions are extracted from the high-resolution polarization measurements shown in Figs. 1d-1g. In agreement with previous ARPES studies\cite{Ye2014,Borisenko2010}, we find that the orbital components of the $\alpha$, $\alpha$', $\beta$ and $\delta$ bands are relatively pure and mainly composed of $d_{xz/yz}$, $d_{xz/yz}$, $d_{xy}$ and $d_{xz/yz}$ orbitals, respectively. The $\gamma$ band however is a mix of the $d_{xy}$ and $d_{xz/yz}$ orbitals, as it can be clearly seen from both the $\sigma$-pol and $\pi$-pol geometries. We note that while the $t_{2g}$ orbitals ($d_{xy}$, $d_{xz}$, $d_{yz}$) have the largest contributions to the density-of-states near $E_{F}$, all bands are slightly mixed with the $e_{g}$ ($d_{x^{2}-y^{2}}$, $d_{z^{2}}$) and $p$ orbitals except at the $\Gamma$ and M high-symmetry points.
%
\begin{figure}[tb]
\includegraphics[width=\columnwidth]{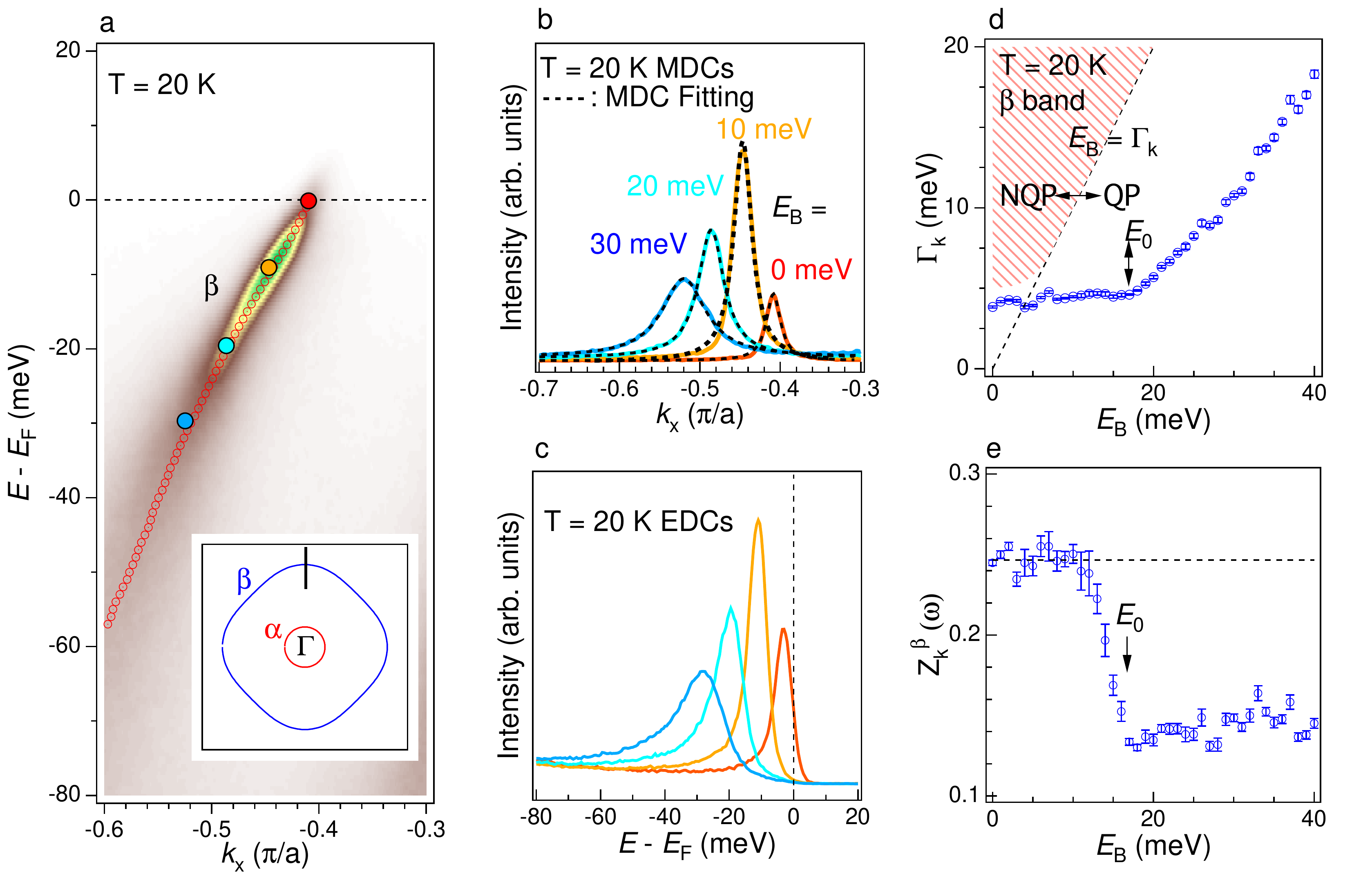}
\caption{ (color online)(a) ARPES intensity plot of the $\beta$ band at 20 K. E/MDCs that are crossing the coloured circles in (a) are shown in (b) and (c). Black dashed curves in (b) are single Lorentz function fittings. d, Extracted energy-dependent scattering rates, Im$\Sigma(\omega)$, are represented by blue circles . The red shaded area represents non-quasi-particle regime. The area does not go to zero because of the temperature correction, $\pi k_{B}T$, to the self-energy. e, Extracted energy-dependent renormalization factor of the $\beta$ band. $E_{0}$ at 16 meV shown in (d) and (e) is corresponding to the energy of the spectral function anomaly\cite{Supp}. Error bars are determined by the standard deviation of the fitting parameters.}
\label{Fig2}
\end{figure}

To prove the existence of well-defined QPs, we look at the well-isolated $\beta$ band, which has the largest band renormalization factor (about 4) and effective mass (about 8 $m_{e}$, where $m_{e}$ is the free electron mass) among all five bands\cite{Ferber2012,Supp}. Figure 2a shows the normal state ARPES intensity plot of the $\beta$ band at 20 K. Selected energy/momentum distribution curves (E/MDCs) of the $\beta$ band shown in Figs. 2b and 2c are labelled by the colored circles in Fig. 2a. Following the common practice\cite{Johnson2001,Valla1999,Supp}, we extracted the energy-dependent scattering rates $\Gamma(k,\omega)$ and the renormalization factor Z(k,$\omega$) of the $\beta$ band, and plotted them in Figs. 2d and 2e, respectively. The extracted scattering rate near $E_{F}$ is dominated by the thermal broadening of the QP lifetime convoluted with our experimental energy and momentum resolutions, and lays deep in the Landau's QP regime, where Im$\Sigma(k,\omega)<\omega$, hence proving the existence of well-defined QP in LiFeAs. The red shaded area represents non-quasi-particle regime. The area does not go to zero because of the temperature correction, $\pi k_{B}T$, to the self-energy. The 16 meV anomalies observed in both $\Gamma(k,\omega)$ and Z(k,$\omega$) are likely induced by electron-boson couplings with negligible contributions of the  antiferromagnetic spin-resonance\cite{Supp}.

%
\begin{figure}[tb]
\includegraphics[width=\columnwidth]{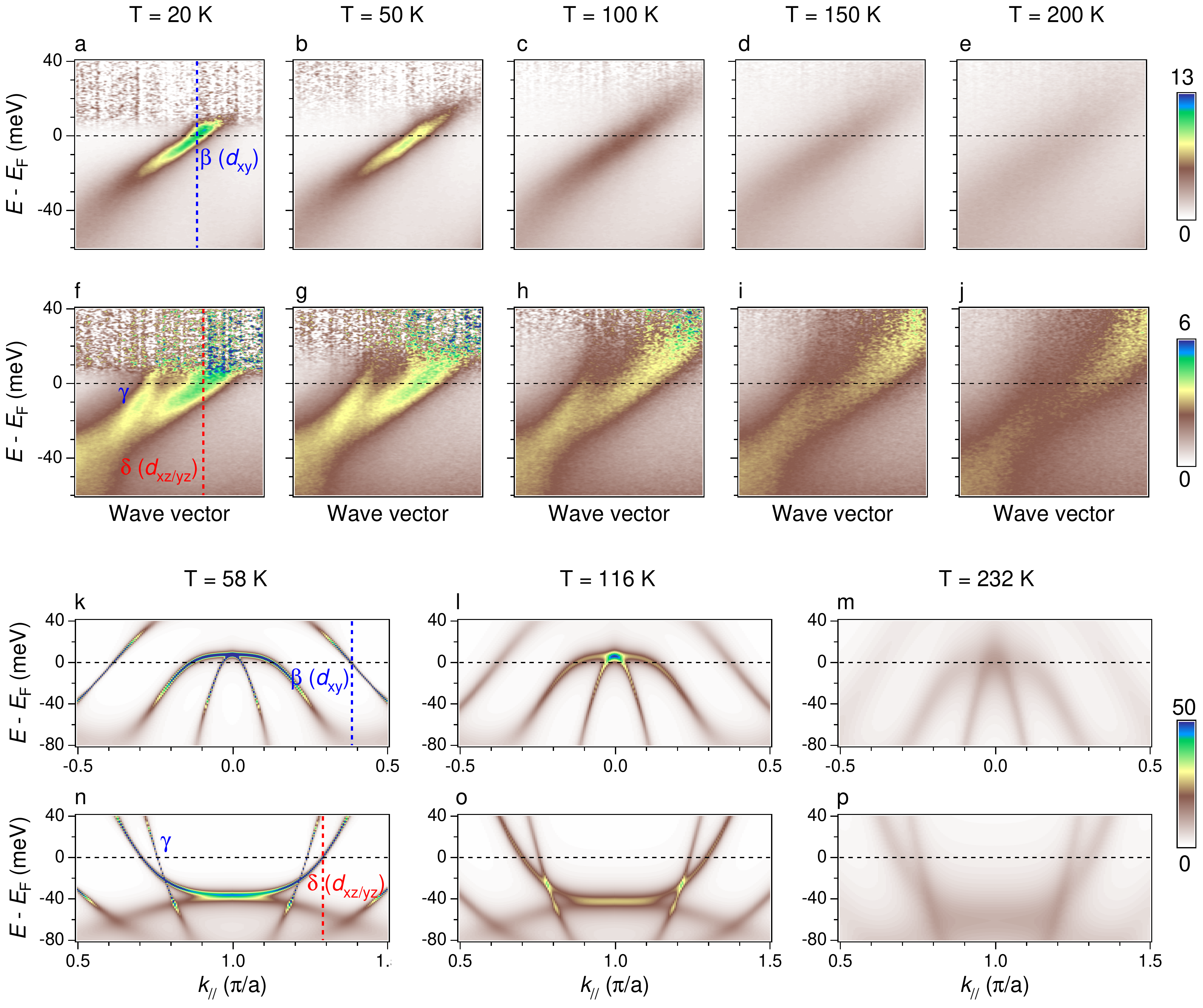}
\caption{ (color online) (a)-(e) ARPES intensity plots of the $\beta$ band at 30 K, 50 K, 100 K, 150 K and 200 K, respectively. (f)-(j) ARPES intensity plots of the $\delta$ and $\gamma$ bands at 30 K, 50 K, 100 K, 150 K and 200 K, respectively. All spectra are divided by the Fermi-Dirac function convoluted with the system resolution. (k)-(p), DFT+DMFT calculated momentum- and energy-resolved spectral function at 58 K, 116 K and 232 K. (k)-(m) and (n)-(p) are corresponding to hole bands and electron bands, respectively.}
\label{Fig3}
\end{figure}
Now we turn to the temperature dependence of the QPs. Figures 3a-3e show ARPES intensity plots of the $\beta$ band at 30 K, 50 K, 100 K, 150 K and 200 K, respectively. Figures 3f-3j show the temperature evolution of the $\gamma$ and $\delta$ bands. To reveal the electronic states above $E_{F}$, all spectra are divided by the Fermi-Dirac function convoluted with the system resolution. As shown in Figs. 3a-3e, the $\beta$ band, which is mainly composed of $d_{xy}$ orbital character, dramatically loses intensity and is nearly invisible at 200 K, while the $\delta$ band, which is mainly composed of $d_{xz/yz}$ orbital, becomes broader and its intensity remains relatively strong even at 200 K. This orbital dependent intensity loss is consistent with previous report on the same material, where the drop of peak intensity on $\beta$ band is much faster than it is on the $\alpha$ and $\alpha$' band, which are mainly composed of $d_{xz/yz}$ orbital\cite{Supp,Miao2014}. In Fig. 3, we show the DFT+DMFT calculated hole bands (Figs. 3k-3m) and electron bands (Figs. 3n-3p) at several temperatures. The overall momentum and energy resolved spectra agree quite well with experimental measurements without any adjustment such as band renormalization and shift, which are usually needed for the DFT band structure, validating the DFT+DMFT approach. It is also evident that the DFT+DMFT intensity of the $\beta$ band with $d_{xy}$ orbital is substantially weaker than the $d_{xz/yz}$ bands at 232 K\cite{Supp}, which is consistent with the experimental observations\cite{Supp}. 

%
%
\begin{figure*}[tb]
\includegraphics[width=16 cm]{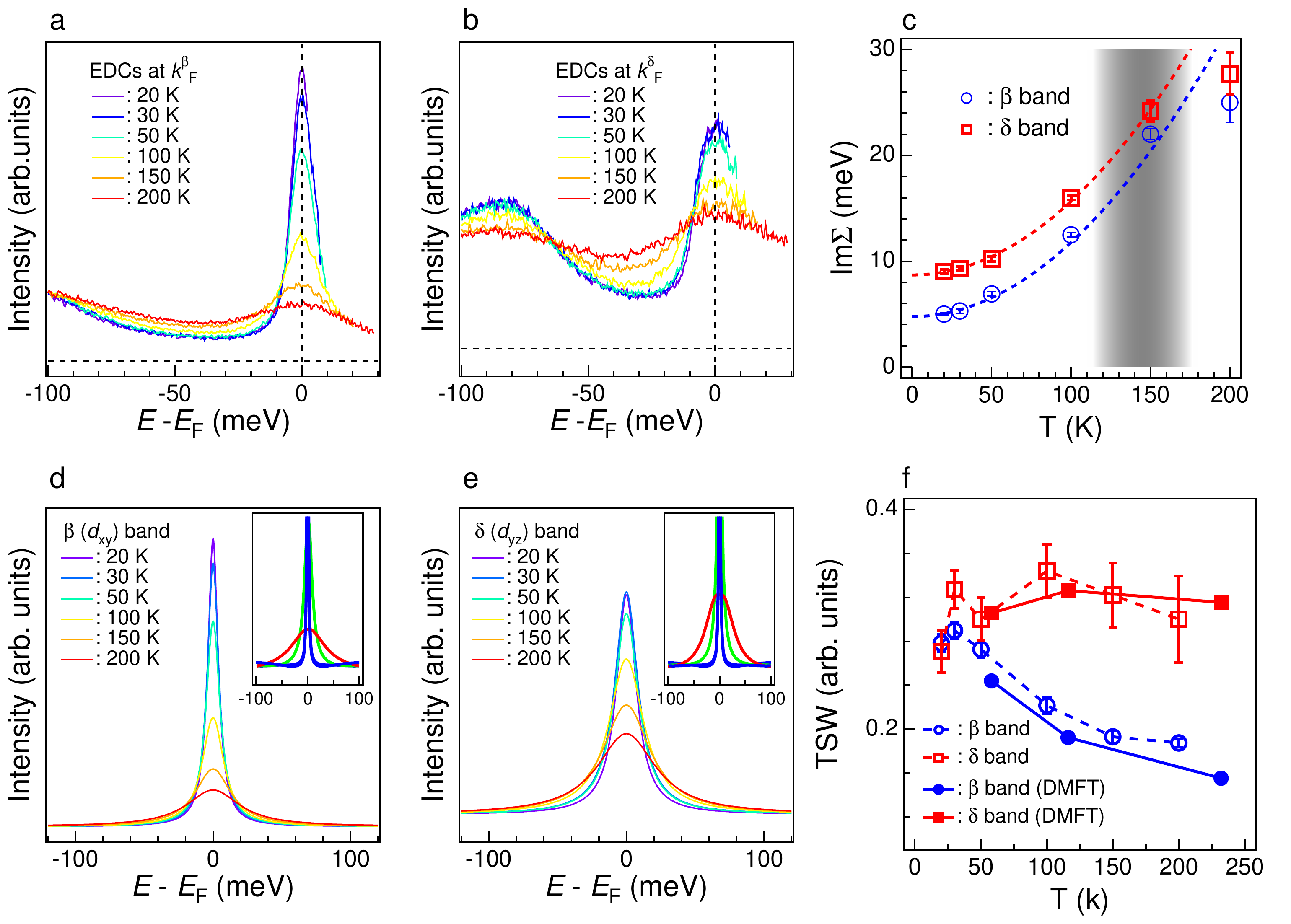}
\caption{ (color online) (a) and (b) EDCs of the $\beta$ and $\delta$ bands from 20 K to 200 K. All curves are fitted by the QP spectral function plus a polynomial background\cite{Supp}. The horizontal dashed lines represent zero intensity. c, Extracted temperature-dependent QP scattering rates. The grey shaded background represents the maximum region of the derivative of the resistivity curve. (d) and (e), Extracted QP peaks at $k_{F}^{\beta}$ and $k_{F}^{\delta}$, respectively. The insets of (d) and (e) are DFT+DMFT calculated spectral functions at $k_{F}^{\beta}$ and $k_{F}^{\delta}$, respectively. To get the total QP spectral weight, we integrate the extracted and calculated spectral functions shown in (d) and (e) and plot the results in (f). Error bars shown in (c) and (f) are determined by the standard deviation of the fitting parameters. }
\label{Fig4}
\end{figure*}

In order to quantitatively compare the difference between the $\beta$ band and the $\delta$ band, we analyze the measured EDCs at $k_{F}^{\beta}$ and $k_{F}^{\delta}$, which are marked by blue and red lines in Figs. 3a and 3f, respectively. In Figs. 4a and 4b, we show the EDCs of the $\beta$ band and the $\delta$ band from 20 K to 200 K. All curves are fitted by the QP spectral function plus a polynomial background and the extracted QP peaks of the $\beta$ and the $\delta$ bands are plotted in Figs. 4d and 4e, respectively\cite{Supp}. The temperature-dependent QP scattering rates are extracted and plotted in Fig. 4c. The grey shaded background represents the coherence-incoherence crossover regime where the derivative of the resistivity curve reaches a maximum and starts to drop down\cite{Kemper2011}. Interestingly, in agreement with a recent study\cite{Brouet2016}, we find that the QP scattering rates on both the $\beta$ band and the $\delta$ band also severely deviate from their low-temperature $T$-quadratic behavior near this temperature, indicating that the saturation of resistivity is intimately connected to the high-temperature QP scattering rate. 

Although the scattering rates of the $\beta$ band and the $\delta$ band show similar temperature evolutions, we find that the total spectral weight (TSW) of the $\beta$ and $\delta$ bands have different behaviors at high temperature. To extract the spectral weight (SW), we integrate the extracted and DFT+DMFT calculated QP spectral functions shown in Figs. 4d and 4e\cite{Supp} and plot the integrated SW of the $\beta$ and $\delta$ bands in Fig. 4f. Both the experimental data and the theoretical calculations show a nearly conserved SW on the $\delta$ band up to 200 K, and a dramatically reduced SW on the $\gamma$ band at high temperature\cite{Supp}. Indeed, the intensity change of the $\alpha$ and $\alpha$' is similar to the $\delta$ band and much slower than the $\beta$ band with increasing temperature\cite{Miao2014,Supp}, further proving the change of SW is orbital dependent.

This orbital-dependent SW reduction with elevated temperature is fully consistent with the Hund's metal picture where an orbital-differentiated coherence-incoherence crossover occurs at different temperatures due to the strong Hund's rule coupling\cite{Yin2012,Mravlje2011}. This is further supported by a recent DMFT plus numerical renormalization group study confirming that the iron pnictides are Fermi liquids at low temperature and the orbital-differentiated coherence-incoherence crossover is driven by a Kondo-type screening with the Kondo temperature determined by the strength of Hund's coupling\cite{Stadler2015}. In the Hund's metal point of view, both iron pnictides and iron chalcogenides have Hund-differentiated coherence-incoherence crossover. Indeed, previous studies\cite{Yi2013,Yi2015} show that both FeTe and K$_{x}$Fe$_{2-y}$Se$_{2}$ exhibit a similar orbital-differentiated coherence-incoherence crossover, with the $d_{xy}$ orbital having the lowest coherent temperature. Therefore, our work proves that the orbital-differentiated coherence-incoherence crossover is a generic feature of the FeSCs. Moreover, this phenomenon is not limited to the FeSCs but is a common feature of all Hund's metals\cite{Mravlje2011}, such as the ruthenates.

Previously, it was also proposed that the FeSCs are in close proximity to a Mott state\cite{Medici2014,Yu2011}. The orbital-dependent SW reduction upon increasing temperature was interpreted as the materials undergo an orbital-selective Mott transition with the $d_{xy}$ orbital being the insulating orbital. Although in this picture the SW will decrease if the system goes towards the orbital-selective Mott state by tuning some parameters, the hybridization of the insulating orbital ($d_{xy}$) with all other orbitals ($d_{xz/yz}$) must vanish to achieve an orbital-selective Mott state\cite{Supp}. However, both LDA calculations and experimental data clearly show hybridization between the $d_{xy}$ orbital and the $d_{xz/yz}$ orbital in LiFeAs is of about 150-300 meV\cite{Borisenko2010, Brouet2016}. This suggests that LiFeAs is far away from the proposed orbital-selective Mott state and reinstates the Hund's coupling induced orbital-differentiated coherence-incoherence crossover viewpoint, in which the reduced spectral weight upon increasing temperature dissolves into an incoherent background with a finite density-of-states at the Fermi level at high temperature due to the strong Hund's coupling $J_{H}$ and the Kondo-screening mechanism, which leads to the decoupling of the orbital and spin degrees of freedom at high temperature. Our results, together with previous studies on the iron-chacolgenides, show that the orbital-differentiated coherence-incoherence crossover is a phenomenon ubiquitous to all FeSCs, and more generally to all Hund's metals whose physics is mainly controlled by the strength of the Hund's coupling.

%
%

We thank J. Schmalian, Z. Wang, T. Valla, P. D. Johnson and W. L. Zhang for useful discussions. The experimental work was supported by grants from CAS (XDB07000000, 112111KYS820150017), MOST (2011CBA001000, 2013CB921700 and 2015CB921301), NSFC (11234014, 11220101003 and 11274362). Theoretical work was supported by NSF-DMR 1308141 (Z. P. Y. and G. K.) and 1405303 (K. H.). H. M. was supported by the Center for Emergent Superconductivity, an Energy Frontier Research Center funded by the U.S. DOE, Office of Basic Energy Sciences. Z. P. Y acknowledges the start-up fund of Beijing Normal University.
\bibliography{biblio}

\begin{thebibliography}{27}%
\makeatletter
\providecommand \@ifxundefined [1]{%
 \@ifx{#1\undefined}
}%
\providecommand \@ifnum [1]{%
 \ifnum #1\expandafter \@firstoftwo
 \else \expandafter \@secondoftwo
 \fi
}%
\providecommand \@ifx [1]{%
 \ifx #1\expandafter \@firstoftwo
 \else \expandafter \@secondoftwo
 \fi
}%
\providecommand \natexlab [1]{#1}%
\providecommand \enquote  [1]{``#1''}%
\providecommand \bibnamefont  [1]{#1}%
\providecommand \bibfnamefont [1]{#1}%
\providecommand \citenamefont [1]{#1}%
\providecommand \href@noop [0]{\@secondoftwo}%
\providecommand \href [0]{\begingroup \@sanitize@url \@href}%
\providecommand \@href[1]{\@@startlink{#1}\@@href}%
\providecommand \@@href[1]{\endgroup#1\@@endlink}%
\providecommand \@sanitize@url [0]{\catcode `\\12\catcode `\$12\catcode
  `\&12\catcode `\#12\catcode `\^12\catcode `\_12\catcode `\%12\relax}%
\providecommand \@@startlink[1]{}%
\providecommand \@@endlink[0]{}%
\providecommand \url  [0]{\begingroup\@sanitize@url \@url }%
\providecommand \@url [1]{\endgroup\@href {#1}{\urlprefix }}%
\providecommand \urlprefix  [0]{URL }%
\providecommand \Eprint [0]{\href }%
\providecommand \doibase [0]{http://dx.doi.org/}%
\providecommand \selectlanguage [0]{\@gobble}%
\providecommand \bibinfo  [0]{\@secondoftwo}%
\providecommand \bibfield  [0]{\@secondoftwo}%
\providecommand \translation [1]{[#1]}%
\providecommand \BibitemOpen [0]{}%
\providecommand \bibitemStop [0]{}%
\providecommand \bibitemNoStop [0]{.\EOS\space}%
\providecommand \EOS [0]{\spacefactor3000\relax}%
\providecommand \BibitemShut  [1]{\csname bibitem#1\endcsname}%
\let\auto@bib@innerbib\@empty
\bibitem [{\citenamefont {Johnson}\ \emph {et~al.}(2001)\citenamefont
  {Johnson}, \citenamefont {Valla}, \citenamefont {Fedorov}, \citenamefont
  {Yusof}, \citenamefont {Wells}, \citenamefont {Li}, \citenamefont
  {Moodenbaugh}, \citenamefont {Gu}, \citenamefont {Koshizuka}, \citenamefont
  {Kendziora}, \citenamefont {Jian},\ and\ \citenamefont
  {Hinks}}]{Johnson2001}%
  \BibitemOpen
  \bibfield  {author} {\bibinfo {author} {\bibfnamefont {P.~D.}\ \bibnamefont
  {Johnson}}, \bibinfo {author} {\bibfnamefont {T.}~\bibnamefont {Valla}},
  \bibinfo {author} {\bibfnamefont {A.~V.}\ \bibnamefont {Fedorov}}, \bibinfo
  {author} {\bibfnamefont {Z.}~\bibnamefont {Yusof}}, \bibinfo {author}
  {\bibfnamefont {B.~O.}\ \bibnamefont {Wells}}, \bibinfo {author}
  {\bibfnamefont {Q.}~\bibnamefont {Li}}, \bibinfo {author} {\bibfnamefont
  {A.~R.}\ \bibnamefont {Moodenbaugh}}, \bibinfo {author} {\bibfnamefont
  {G.~D.}\ \bibnamefont {Gu}}, \bibinfo {author} {\bibfnamefont
  {N.}~\bibnamefont {Koshizuka}}, \bibinfo {author} {\bibfnamefont
  {C.}~\bibnamefont {Kendziora}}, \bibinfo {author} {\bibfnamefont
  {S.}~\bibnamefont {Jian}}, \ and\ \bibinfo {author} {\bibfnamefont {D.~G.}\
  \bibnamefont {Hinks}},\ }\href {\doibase 10.1103/PhysRevLett.87.177007}
  {\bibfield  {journal} {\bibinfo  {journal} {Phys. Rev. Lett.}\ }\textbf
  {\bibinfo {volume} {87}},\ \bibinfo {pages} {177007} (\bibinfo {year}
  {2001})}\BibitemShut {NoStop}%
\bibitem [{\citenamefont {Damascelli}\ \emph {et~al.}(2003)\citenamefont
  {Damascelli}, \citenamefont {Hussain},\ and\ \citenamefont
  {Shen}}]{Shen2003}%
  \BibitemOpen
  \bibfield  {author} {\bibinfo {author} {\bibfnamefont {A.}~\bibnamefont
  {Damascelli}}, \bibinfo {author} {\bibfnamefont {Z.}~\bibnamefont {Hussain}},
  \ and\ \bibinfo {author} {\bibfnamefont {Z.-X.}\ \bibnamefont {Shen}},\
  }\href {\doibase 10.1103/RevModPhys.75.473} {\bibfield  {journal} {\bibinfo
  {journal} {Rev. Mod. Phys.}\ }\textbf {\bibinfo {volume} {75}},\ \bibinfo
  {pages} {473} (\bibinfo {year} {2003})}\BibitemShut {NoStop}%
\bibitem [{\citenamefont {Valla}\ \emph {et~al.}(1999)\citenamefont {Valla},
  \citenamefont {Fedorov}, \citenamefont {Johnson},\ and\ \citenamefont
  {Hulbert}}]{Valla1999}%
  \BibitemOpen
  \bibfield  {author} {\bibinfo {author} {\bibfnamefont {T.}~\bibnamefont
  {Valla}}, \bibinfo {author} {\bibfnamefont {A.~V.}\ \bibnamefont {Fedorov}},
  \bibinfo {author} {\bibfnamefont {P.~D.}\ \bibnamefont {Johnson}}, \ and\
  \bibinfo {author} {\bibfnamefont {S.~L.}\ \bibnamefont {Hulbert}},\ }\href
  {\doibase 10.1103/PhysRevLett.83.2085} {\bibfield  {journal} {\bibinfo
  {journal} {Phys. Rev. Lett.}\ }\textbf {\bibinfo {volume} {83}},\ \bibinfo
  {pages} {2085} (\bibinfo {year} {1999})}\BibitemShut {NoStop}%
\bibitem [{\citenamefont {Richard}\ \emph {et~al.}(2009)\citenamefont
  {Richard}, \citenamefont {Sato}, \citenamefont {Nakayama}, \citenamefont
  {Souma}, \citenamefont {Takahashi}, \citenamefont {Xu}, \citenamefont {Chen},
  \citenamefont {Luo}, \citenamefont {Wang},\ and\ \citenamefont
  {Ding}}]{Richard2009}%
  \BibitemOpen
  \bibfield  {author} {\bibinfo {author} {\bibfnamefont {P.}~\bibnamefont
  {Richard}}, \bibinfo {author} {\bibfnamefont {T.}~\bibnamefont {Sato}},
  \bibinfo {author} {\bibfnamefont {K.}~\bibnamefont {Nakayama}}, \bibinfo
  {author} {\bibfnamefont {S.}~\bibnamefont {Souma}}, \bibinfo {author}
  {\bibfnamefont {T.}~\bibnamefont {Takahashi}}, \bibinfo {author}
  {\bibfnamefont {Y.-M.}\ \bibnamefont {Xu}}, \bibinfo {author} {\bibfnamefont
  {G.~F.}\ \bibnamefont {Chen}}, \bibinfo {author} {\bibfnamefont {J.~L.}\
  \bibnamefont {Luo}}, \bibinfo {author} {\bibfnamefont {N.~L.}\ \bibnamefont
  {Wang}}, \ and\ \bibinfo {author} {\bibfnamefont {H.}~\bibnamefont {Ding}},\
  }\href {\doibase 10.1103/PhysRevLett.102.047003} {\bibfield  {journal}
  {\bibinfo  {journal} {Phys. Rev. Lett.}\ }\textbf {\bibinfo {volume} {102}},\
  \bibinfo {pages} {047003} (\bibinfo {year} {2009})}\BibitemShut {NoStop}%
\bibitem [{\citenamefont {Norman}\ \emph {et~al.}(1997)\citenamefont {Norman},
  \citenamefont {Ding}, \citenamefont {Campuzano}, \citenamefont {Takeuchi},
  \citenamefont {Randeria}, \citenamefont {Yokoya}, \citenamefont {Takahashi},
  \citenamefont {Mochiku},\ and\ \citenamefont {Kadowaki}}]{Norman1997}%
  \BibitemOpen
  \bibfield  {author} {\bibinfo {author} {\bibfnamefont {M.~R.}\ \bibnamefont
  {Norman}}, \bibinfo {author} {\bibfnamefont {H.}~\bibnamefont {Ding}},
  \bibinfo {author} {\bibfnamefont {J.~C.}\ \bibnamefont {Campuzano}}, \bibinfo
  {author} {\bibfnamefont {T.}~\bibnamefont {Takeuchi}}, \bibinfo {author}
  {\bibfnamefont {M.}~\bibnamefont {Randeria}}, \bibinfo {author}
  {\bibfnamefont {T.}~\bibnamefont {Yokoya}}, \bibinfo {author} {\bibfnamefont
  {T.}~\bibnamefont {Takahashi}}, \bibinfo {author} {\bibfnamefont
  {T.}~\bibnamefont {Mochiku}}, \ and\ \bibinfo {author} {\bibfnamefont
  {K.}~\bibnamefont {Kadowaki}},\ }\href {\doibase 10.1103/PhysRevLett.79.3506}
  {\bibfield  {journal} {\bibinfo  {journal} {Phys. Rev. Lett.}\ }\textbf
  {\bibinfo {volume} {79}},\ \bibinfo {pages} {3506} (\bibinfo {year}
  {1997})}\BibitemShut {NoStop}%
\bibitem [{\citenamefont {Norman}\ \emph {et~al.}(1998)\citenamefont {Norman},
  \citenamefont {Randeria}, \citenamefont {Ding},\ and\ \citenamefont
  {Campuzano}}]{Norman1998}%
  \BibitemOpen
  \bibfield  {author} {\bibinfo {author} {\bibfnamefont {M.~R.}\ \bibnamefont
  {Norman}}, \bibinfo {author} {\bibfnamefont {M.}~\bibnamefont {Randeria}},
  \bibinfo {author} {\bibfnamefont {H.}~\bibnamefont {Ding}}, \ and\ \bibinfo
  {author} {\bibfnamefont {J.~C.}\ \bibnamefont {Campuzano}},\ }\href {\doibase
  10.1103/PhysRevB.57.R11093} {\bibfield  {journal} {\bibinfo  {journal} {Phys.
  Rev. B}\ }\textbf {\bibinfo {volume} {57}},\ \bibinfo {pages} {R11093}
  (\bibinfo {year} {1998})}\BibitemShut {NoStop}%
\bibitem [{\citenamefont {Yi}\ \emph {et~al.}(2013)\citenamefont {Yi},
  \citenamefont {Lu}, \citenamefont {Yu}, \citenamefont {Riggs}, \citenamefont
  {Chu}, \citenamefont {Lv}, \citenamefont {Liu}, \citenamefont {Lu},
  \citenamefont {Cui}, \citenamefont {Hashimoto}, \citenamefont {Mo},
  \citenamefont {Hussain}, \citenamefont {Chu}, \citenamefont {Fisher},
  \citenamefont {Si},\ and\ \citenamefont {Shen}}]{Yi2013}%
  \BibitemOpen
  \bibfield  {author} {\bibinfo {author} {\bibfnamefont {M.}~\bibnamefont
  {Yi}}, \bibinfo {author} {\bibfnamefont {D.~H.}\ \bibnamefont {Lu}}, \bibinfo
  {author} {\bibfnamefont {R.}~\bibnamefont {Yu}}, \bibinfo {author}
  {\bibfnamefont {S.~C.}\ \bibnamefont {Riggs}}, \bibinfo {author}
  {\bibfnamefont {J.-H.}\ \bibnamefont {Chu}}, \bibinfo {author} {\bibfnamefont
  {B.}~\bibnamefont {Lv}}, \bibinfo {author} {\bibfnamefont {Z.~K.}\
  \bibnamefont {Liu}}, \bibinfo {author} {\bibfnamefont {M.}~\bibnamefont
  {Lu}}, \bibinfo {author} {\bibfnamefont {Y.-T.}\ \bibnamefont {Cui}},
  \bibinfo {author} {\bibfnamefont {M.}~\bibnamefont {Hashimoto}}, \bibinfo
  {author} {\bibfnamefont {S.-K.}\ \bibnamefont {Mo}}, \bibinfo {author}
  {\bibfnamefont {Z.}~\bibnamefont {Hussain}}, \bibinfo {author} {\bibfnamefont
  {C.~W.}\ \bibnamefont {Chu}}, \bibinfo {author} {\bibfnamefont {I.~R.}\
  \bibnamefont {Fisher}}, \bibinfo {author} {\bibfnamefont {Q.}~\bibnamefont
  {Si}}, \ and\ \bibinfo {author} {\bibfnamefont {Z.-X.}\ \bibnamefont
  {Shen}},\ }\href {\doibase 10.1103/PhysRevLett.110.067003} {\bibfield
  {journal} {\bibinfo  {journal} {Phys. Rev. Lett.}\ }\textbf {\bibinfo
  {volume} {110}},\ \bibinfo {pages} {067003} (\bibinfo {year}
  {2013})}\BibitemShut {NoStop}%
\bibitem [{\citenamefont {Yi}\ \emph {et~al.}(2015)\citenamefont {Yi},
  \citenamefont {Liu}, \citenamefont {Zhang}, \citenamefont {Yu}, \citenamefont
  {Zhu}, \citenamefont {Lee}, \citenamefont {Moore}, \citenamefont {Schmitt},
  \citenamefont {Li}, \citenamefont {Riggs}, \citenamefont {Chu}, \citenamefont
  {Lv}, \citenamefont {Hu}, \citenamefont {Hashimoto}, \citenamefont {Mo},
  \citenamefont {Hussain}, \citenamefont {Mao}, \citenamefont {Chu},
  \citenamefont {Fisher}, \citenamefont {Si}, \citenamefont {Shen},\ and\
  \citenamefont {Lu}}]{Yi2015}%
  \BibitemOpen
  \bibfield  {author} {\bibinfo {author} {\bibfnamefont {M.}~\bibnamefont
  {Yi}}, \bibinfo {author} {\bibfnamefont {Z.-K.}\ \bibnamefont {Liu}},
  \bibinfo {author} {\bibfnamefont {Y.}~\bibnamefont {Zhang}}, \bibinfo
  {author} {\bibfnamefont {R.}~\bibnamefont {Yu}}, \bibinfo {author}
  {\bibfnamefont {J.-X.}\ \bibnamefont {Zhu}}, \bibinfo {author} {\bibfnamefont
  {J.~J.}\ \bibnamefont {Lee}}, \bibinfo {author} {\bibfnamefont {R.~G.}\
  \bibnamefont {Moore}}, \bibinfo {author} {\bibfnamefont {F.~T.}\ \bibnamefont
  {Schmitt}}, \bibinfo {author} {\bibfnamefont {W.}~\bibnamefont {Li}},
  \bibinfo {author} {\bibfnamefont {S.~C.}\ \bibnamefont {Riggs}}, \bibinfo
  {author} {\bibfnamefont {J.-H.}\ \bibnamefont {Chu}}, \bibinfo {author}
  {\bibfnamefont {B.}~\bibnamefont {Lv}}, \bibinfo {author} {\bibfnamefont
  {J.}~\bibnamefont {Hu}}, \bibinfo {author} {\bibfnamefont {M.}~\bibnamefont
  {Hashimoto}}, \bibinfo {author} {\bibfnamefont {S.-K.}\ \bibnamefont {Mo}},
  \bibinfo {author} {\bibfnamefont {Z.}~\bibnamefont {Hussain}}, \bibinfo
  {author} {\bibfnamefont {Z.~Q.}\ \bibnamefont {Mao}}, \bibinfo {author}
  {\bibfnamefont {C.~W.}\ \bibnamefont {Chu}}, \bibinfo {author} {\bibfnamefont
  {I.~R.}\ \bibnamefont {Fisher}}, \bibinfo {author} {\bibfnamefont
  {Q.}~\bibnamefont {Si}}, \bibinfo {author} {\bibfnamefont {Z.-X.}\
  \bibnamefont {Shen}}, \ and\ \bibinfo {author} {\bibfnamefont {D.~H.}\
  \bibnamefont {Lu}},\ }\href {http://dx.doi.org/10.1038/ncomms8777
  10.1038/ncomms8777} {\bibfield  {journal} {\bibinfo  {journal} {Nat Commun}\
  }\textbf {\bibinfo {volume} {6}} (\bibinfo {year} {2015})}\BibitemShut
  {NoStop}%
\bibitem [{\citenamefont {Yu}\ and\ \citenamefont {Si}(2011)}]{Yu2011}%
  \BibitemOpen
  \bibfield  {author} {\bibinfo {author} {\bibfnamefont {R.}~\bibnamefont
  {Yu}}\ and\ \bibinfo {author} {\bibfnamefont {Q.}~\bibnamefont {Si}},\ }\href
  {\doibase 10.1103/PhysRevB.84.235115} {\bibfield  {journal} {\bibinfo
  {journal} {Phys. Rev. B}\ }\textbf {\bibinfo {volume} {84}},\ \bibinfo
  {pages} {235115} (\bibinfo {year} {2011})}\BibitemShut {NoStop}%
\bibitem [{\citenamefont {de' Medici}\ \emph {et~al.}(2014)\citenamefont {de'
  Medici}, \citenamefont {Giovannetti},\ and\ \citenamefont
  {Capone}}]{Medici2014}%
  \BibitemOpen
  \bibfield  {author} {\bibinfo {author} {\bibfnamefont {L.}~\bibnamefont {de'
  Medici}}, \bibinfo {author} {\bibfnamefont {G.}~\bibnamefont {Giovannetti}},
  \ and\ \bibinfo {author} {\bibfnamefont {M.}~\bibnamefont {Capone}},\ }\href
  {\doibase 10.1103/PhysRevLett.112.177001} {\bibfield  {journal} {\bibinfo
  {journal} {Phys. Rev. Lett.}\ }\textbf {\bibinfo {volume} {112}},\ \bibinfo
  {pages} {177001} (\bibinfo {year} {2014})}\BibitemShut {NoStop}%
\bibitem [{\citenamefont {Yin}\ \emph {et~al.}(2011{\natexlab{a}})\citenamefont
  {Yin}, \citenamefont {Haule},\ and\ \citenamefont {Kotliar}}]{Yin2011a}%
  \BibitemOpen
  \bibfield  {author} {\bibinfo {author} {\bibfnamefont {Z.~P.}\ \bibnamefont
  {Yin}}, \bibinfo {author} {\bibfnamefont {K.}~\bibnamefont {Haule}}, \ and\
  \bibinfo {author} {\bibfnamefont {G.}~\bibnamefont {Kotliar}},\ }\href
  {http://dx.doi.org/10.1038/nmat3120
  http://www.nature.com/nmat/journal/v10/n12/abs/nmat3120.html{\#}supplementary-information}
  {\bibfield  {journal} {\bibinfo  {journal} {Nat Mater}\ }\textbf {\bibinfo
  {volume} {10}},\ \bibinfo {pages} {932} (\bibinfo {year}
  {2011}{\natexlab{a}})}\BibitemShut {NoStop}%
\bibitem [{\citenamefont {Yin}\ \emph {et~al.}(2011{\natexlab{b}})\citenamefont
  {Yin}, \citenamefont {Haule},\ and\ \citenamefont {Kotliar}}]{Yin2011b}%
  \BibitemOpen
  \bibfield  {author} {\bibinfo {author} {\bibfnamefont {Z.~P.}\ \bibnamefont
  {Yin}}, \bibinfo {author} {\bibfnamefont {K.}~\bibnamefont {Haule}}, \ and\
  \bibinfo {author} {\bibfnamefont {G.}~\bibnamefont {Kotliar}},\ }\href
  {http://dx.doi.org/10.1038/nphys1923
  http://www.nature.com/nphys/journal/v7/n4/abs/nphys1923.html{\#}supplementary-information}
  {\bibfield  {journal} {\bibinfo  {journal} {Nat Phys}\ }\textbf {\bibinfo
  {volume} {7}},\ \bibinfo {pages} {294} (\bibinfo {year}
  {2011}{\natexlab{b}})}\BibitemShut {NoStop}%
\bibitem [{\citenamefont {Miao}\ \emph {et~al.}(2014)\citenamefont {Miao},
  \citenamefont {Wang}, \citenamefont {Richard}, \citenamefont {Wu},
  \citenamefont {Ma}, \citenamefont {Qian}, \citenamefont {Xing}, \citenamefont
  {Wang}, \citenamefont {Jin}, \citenamefont {Chou}, \citenamefont {Wang},
  \citenamefont {Ku},\ and\ \citenamefont {Ding}}]{Miao2014}%
  \BibitemOpen
  \bibfield  {author} {\bibinfo {author} {\bibfnamefont {H.}~\bibnamefont
  {Miao}}, \bibinfo {author} {\bibfnamefont {L.-M.}\ \bibnamefont {Wang}},
  \bibinfo {author} {\bibfnamefont {P.}~\bibnamefont {Richard}}, \bibinfo
  {author} {\bibfnamefont {S.-F.}\ \bibnamefont {Wu}}, \bibinfo {author}
  {\bibfnamefont {J.}~\bibnamefont {Ma}}, \bibinfo {author} {\bibfnamefont
  {T.}~\bibnamefont {Qian}}, \bibinfo {author} {\bibfnamefont {L.-Y.}\
  \bibnamefont {Xing}}, \bibinfo {author} {\bibfnamefont {X.-C.}\ \bibnamefont
  {Wang}}, \bibinfo {author} {\bibfnamefont {C.-Q.}\ \bibnamefont {Jin}},
  \bibinfo {author} {\bibfnamefont {C.-P.}\ \bibnamefont {Chou}}, \bibinfo
  {author} {\bibfnamefont {Z.}~\bibnamefont {Wang}}, \bibinfo {author}
  {\bibfnamefont {W.}~\bibnamefont {Ku}}, \ and\ \bibinfo {author}
  {\bibfnamefont {H.}~\bibnamefont {Ding}},\ }\href {\doibase
  10.1103/PhysRevB.89.220503} {\bibfield  {journal} {\bibinfo  {journal} {Phys.
  Rev. B}\ }\textbf {\bibinfo {volume} {89}},\ \bibinfo {pages} {220503}
  (\bibinfo {year} {2014})}\BibitemShut {NoStop}%
\bibitem [{\citenamefont {Haule}\ \emph {et~al.}(2010)\citenamefont {Haule},
  \citenamefont {Yee},\ and\ \citenamefont {Kim}}]{Haule2010}%
  \BibitemOpen
  \bibfield  {author} {\bibinfo {author} {\bibfnamefont {K.}~\bibnamefont
  {Haule}}, \bibinfo {author} {\bibfnamefont {C.-H.}\ \bibnamefont {Yee}}, \
  and\ \bibinfo {author} {\bibfnamefont {K.}~\bibnamefont {Kim}},\ }\href
  {\doibase 10.1103/PhysRevB.81.195107} {\bibfield  {journal} {\bibinfo
  {journal} {Phys. Rev. B}\ }\textbf {\bibinfo {volume} {81}},\ \bibinfo
  {pages} {195107} (\bibinfo {year} {2010})}\BibitemShut {NoStop}%
\bibitem [{\citenamefont {Yin}\ \emph {et~al.}(2014)\citenamefont {Yin},
  \citenamefont {Haule},\ and\ \citenamefont {Kotliar}}]{Yin2014}%
  \BibitemOpen
  \bibfield  {author} {\bibinfo {author} {\bibfnamefont {Z.~P.}\ \bibnamefont
  {Yin}}, \bibinfo {author} {\bibfnamefont {K.}~\bibnamefont {Haule}}, \ and\
  \bibinfo {author} {\bibfnamefont {G.}~\bibnamefont {Kotliar}},\ }\href
  {http://dx.doi.org/10.1038/nphys3116 10.1038/nphys3116
  http://www.nature.com/nphys/journal/v10/n11/abs/nphys3116.html{\#}supplementary-information}
  {\bibfield  {journal} {\bibinfo  {journal} {Nat Phys}\ }\textbf {\bibinfo
  {volume} {10}},\ \bibinfo {pages} {845} (\bibinfo {year} {2014})}\BibitemShut
  {NoStop}%
\bibitem [{\citenamefont {Miao}\ \emph {et~al.}(2015)\citenamefont {Miao},
  \citenamefont {Qian}, \citenamefont {Shi}, \citenamefont {Richard},
  \citenamefont {Kim}, \citenamefont {Hoesch}, \citenamefont {Xing},
  \citenamefont {Wang}, \citenamefont {Jin}, \citenamefont {Hu},\ and\
  \citenamefont {Ding}}]{Miao2015}%
  \BibitemOpen
  \bibfield  {author} {\bibinfo {author} {\bibfnamefont {H.}~\bibnamefont
  {Miao}}, \bibinfo {author} {\bibfnamefont {T.}~\bibnamefont {Qian}}, \bibinfo
  {author} {\bibfnamefont {X.}~\bibnamefont {Shi}}, \bibinfo {author}
  {\bibfnamefont {P.}~\bibnamefont {Richard}}, \bibinfo {author} {\bibfnamefont
  {T.~K.}\ \bibnamefont {Kim}}, \bibinfo {author} {\bibfnamefont
  {M.}~\bibnamefont {Hoesch}}, \bibinfo {author} {\bibfnamefont {L.~Y.}\
  \bibnamefont {Xing}}, \bibinfo {author} {\bibfnamefont {X.-C.}\ \bibnamefont
  {Wang}}, \bibinfo {author} {\bibfnamefont {C.-Q.}\ \bibnamefont {Jin}},
  \bibinfo {author} {\bibfnamefont {J.-P.}\ \bibnamefont {Hu}}, \ and\ \bibinfo
  {author} {\bibfnamefont {H.}~\bibnamefont {Ding}},\ }\href
  {http://dx.doi.org/10.1038/ncomms7056 10.1038/ncomms7056} {\bibfield
  {journal} {\bibinfo  {journal} {Nat Commun}\ }\textbf {\bibinfo {volume} {6}}
  (\bibinfo {year} {2015})}\BibitemShut {NoStop}%
\bibitem [{\citenamefont {Ye}\ \emph {et~al.}(2014)\citenamefont {Ye},
  \citenamefont {Zhang}, \citenamefont {Chen}, \citenamefont {Xu},
  \citenamefont {Jiang}, \citenamefont {Niu}, \citenamefont {Wen},
  \citenamefont {Xing}, \citenamefont {Wang}, \citenamefont {Jin},
  \citenamefont {Xie},\ and\ \citenamefont {Feng}}]{Ye2014}%
  \BibitemOpen
  \bibfield  {author} {\bibinfo {author} {\bibfnamefont {Z.~R.}\ \bibnamefont
  {Ye}}, \bibinfo {author} {\bibfnamefont {Y.}~\bibnamefont {Zhang}}, \bibinfo
  {author} {\bibfnamefont {F.}~\bibnamefont {Chen}}, \bibinfo {author}
  {\bibfnamefont {M.}~\bibnamefont {Xu}}, \bibinfo {author} {\bibfnamefont
  {J.}~\bibnamefont {Jiang}}, \bibinfo {author} {\bibfnamefont {X.~H.}\
  \bibnamefont {Niu}}, \bibinfo {author} {\bibfnamefont {C.~H.~P.}\
  \bibnamefont {Wen}}, \bibinfo {author} {\bibfnamefont {L.~Y.}\ \bibnamefont
  {Xing}}, \bibinfo {author} {\bibfnamefont {X.~C.}\ \bibnamefont {Wang}},
  \bibinfo {author} {\bibfnamefont {C.~Q.}\ \bibnamefont {Jin}}, \bibinfo
  {author} {\bibfnamefont {B.~P.}\ \bibnamefont {Xie}}, \ and\ \bibinfo
  {author} {\bibfnamefont {D.~L.}\ \bibnamefont {Feng}},\ }\href {\doibase
  10.1103/PhysRevX.4.031041} {\bibfield  {journal} {\bibinfo  {journal} {Phys.
  Rev. X}\ }\textbf {\bibinfo {volume} {4}},\ \bibinfo {pages} {031041}
  (\bibinfo {year} {2014})}\BibitemShut {NoStop}%
\bibitem [{\citenamefont {Dai}\ \emph {et~al.}(2015)\citenamefont {Dai},
  \citenamefont {Miao}, \citenamefont {Xing}, \citenamefont {Wang},
  \citenamefont {Wang}, \citenamefont {Xiao}, \citenamefont {Qian},
  \citenamefont {Richard}, \citenamefont {Qiu}, \citenamefont {Yu},
  \citenamefont {Jin}, \citenamefont {Wang}, \citenamefont {Johnson},
  \citenamefont {Homes},\ and\ \citenamefont {Ding}}]{Dai2015}%
  \BibitemOpen
  \bibfield  {author} {\bibinfo {author} {\bibfnamefont {Y.~M.}\ \bibnamefont
  {Dai}}, \bibinfo {author} {\bibfnamefont {H.}~\bibnamefont {Miao}}, \bibinfo
  {author} {\bibfnamefont {L.~Y.}\ \bibnamefont {Xing}}, \bibinfo {author}
  {\bibfnamefont {X.~C.}\ \bibnamefont {Wang}}, \bibinfo {author}
  {\bibfnamefont {P.~S.}\ \bibnamefont {Wang}}, \bibinfo {author}
  {\bibfnamefont {H.}~\bibnamefont {Xiao}}, \bibinfo {author} {\bibfnamefont
  {T.}~\bibnamefont {Qian}}, \bibinfo {author} {\bibfnamefont {P.}~\bibnamefont
  {Richard}}, \bibinfo {author} {\bibfnamefont {X.~G.}\ \bibnamefont {Qiu}},
  \bibinfo {author} {\bibfnamefont {W.}~\bibnamefont {Yu}}, \bibinfo {author}
  {\bibfnamefont {C.~Q.}\ \bibnamefont {Jin}}, \bibinfo {author} {\bibfnamefont
  {Z.}~\bibnamefont {Wang}}, \bibinfo {author} {\bibfnamefont {P.~D.}\
  \bibnamefont {Johnson}}, \bibinfo {author} {\bibfnamefont {C.~C.}\
  \bibnamefont {Homes}}, \ and\ \bibinfo {author} {\bibfnamefont
  {H.}~\bibnamefont {Ding}},\ }\href {\doibase 10.1103/PhysRevX.5.031035}
  {\bibfield  {journal} {\bibinfo  {journal} {Phys. Rev. X}\ }\textbf {\bibinfo
  {volume} {5}},\ \bibinfo {pages} {031035} (\bibinfo {year}
  {2015})}\BibitemShut {NoStop}%
\bibitem [{\citenamefont {Wang}\ \emph {et~al.}(2012)\citenamefont {Wang},
  \citenamefont {Richard}, \citenamefont {Huang}, \citenamefont {Miao},
  \citenamefont {Cevey}, \citenamefont {Xu}, \citenamefont {Sun}, \citenamefont
  {Qian}, \citenamefont {Xu}, \citenamefont {Shi}, \citenamefont {Hu},
  \citenamefont {Dai},\ and\ \citenamefont {Ding}}]{Wang2012}%
  \BibitemOpen
  \bibfield  {author} {\bibinfo {author} {\bibfnamefont {X.-P.}\ \bibnamefont
  {Wang}}, \bibinfo {author} {\bibfnamefont {P.}~\bibnamefont {Richard}},
  \bibinfo {author} {\bibfnamefont {Y.-B.}\ \bibnamefont {Huang}}, \bibinfo
  {author} {\bibfnamefont {H.}~\bibnamefont {Miao}}, \bibinfo {author}
  {\bibfnamefont {L.}~\bibnamefont {Cevey}}, \bibinfo {author} {\bibfnamefont
  {N.}~\bibnamefont {Xu}}, \bibinfo {author} {\bibfnamefont {Y.-J.}\
  \bibnamefont {Sun}}, \bibinfo {author} {\bibfnamefont {T.}~\bibnamefont
  {Qian}}, \bibinfo {author} {\bibfnamefont {Y.-M.}\ \bibnamefont {Xu}},
  \bibinfo {author} {\bibfnamefont {M.}~\bibnamefont {Shi}}, \bibinfo {author}
  {\bibfnamefont {J.-P.}\ \bibnamefont {Hu}}, \bibinfo {author} {\bibfnamefont
  {X.}~\bibnamefont {Dai}}, \ and\ \bibinfo {author} {\bibfnamefont
  {H.}~\bibnamefont {Ding}},\ }\href {\doibase 10.1103/PhysRevB.85.214518}
  {\bibfield  {journal} {\bibinfo  {journal} {Phys. Rev. B}\ }\textbf {\bibinfo
  {volume} {85}},\ \bibinfo {pages} {214518} (\bibinfo {year}
  {2012})}\BibitemShut {NoStop}%
\bibitem [{\citenamefont {Borisenko}\ \emph {et~al.}(2010)\citenamefont
  {Borisenko}, \citenamefont {Zabolotnyy}, \citenamefont {Evtushinsky},
  \citenamefont {Kim}, \citenamefont {Morozov}, \citenamefont {Yaresko},
  \citenamefont {Kordyuk}, \citenamefont {Behr}, \citenamefont {Vasiliev},
  \citenamefont {Follath},\ and\ \citenamefont {B\"uchner}}]{Borisenko2010}%
  \BibitemOpen
  \bibfield  {author} {\bibinfo {author} {\bibfnamefont {S.~V.}\ \bibnamefont
  {Borisenko}}, \bibinfo {author} {\bibfnamefont {V.~B.}\ \bibnamefont
  {Zabolotnyy}}, \bibinfo {author} {\bibfnamefont {D.~V.}\ \bibnamefont
  {Evtushinsky}}, \bibinfo {author} {\bibfnamefont {T.~K.}\ \bibnamefont
  {Kim}}, \bibinfo {author} {\bibfnamefont {I.~V.}\ \bibnamefont {Morozov}},
  \bibinfo {author} {\bibfnamefont {A.~N.}\ \bibnamefont {Yaresko}}, \bibinfo
  {author} {\bibfnamefont {A.~A.}\ \bibnamefont {Kordyuk}}, \bibinfo {author}
  {\bibfnamefont {G.}~\bibnamefont {Behr}}, \bibinfo {author} {\bibfnamefont
  {A.}~\bibnamefont {Vasiliev}}, \bibinfo {author} {\bibfnamefont
  {R.}~\bibnamefont {Follath}}, \ and\ \bibinfo {author} {\bibfnamefont
  {B.}~\bibnamefont {B\"uchner}},\ }\href {\doibase
  10.1103/PhysRevLett.105.067002} {\bibfield  {journal} {\bibinfo  {journal}
  {Phys. Rev. Lett.}\ }\textbf {\bibinfo {volume} {105}},\ \bibinfo {pages}
  {067002} (\bibinfo {year} {2010})}\BibitemShut {NoStop}%
\bibitem [{Sup()}]{Supp}%
  \BibitemOpen
  \href@noop {} {\bibinfo  {journal} {Supplementary materials}\ }\BibitemShut
  {NoStop}%
\bibitem [{\citenamefont {Ferber}\ \emph {et~al.}(2012)\citenamefont {Ferber},
  \citenamefont {Foyevtsova}, \citenamefont {Valent\'{\i}},\ and\ \citenamefont
  {Jeschke}}]{Ferber2012}%
  \BibitemOpen
\bibfield  {journal} {  }\bibfield  {author} {\bibinfo {author} {\bibfnamefont
  {J.}~\bibnamefont {Ferber}}, \bibinfo {author} {\bibfnamefont
  {K.}~\bibnamefont {Foyevtsova}}, \bibinfo {author} {\bibfnamefont
  {R.}~\bibnamefont {Valent\'{\i}}}, \ and\ \bibinfo {author} {\bibfnamefont
  {H.~O.}\ \bibnamefont {Jeschke}},\ }\href {\doibase
  10.1103/PhysRevB.85.094505} {\bibfield  {journal} {\bibinfo  {journal} {Phys.
  Rev. B}\ }\textbf {\bibinfo {volume} {85}},\ \bibinfo {pages} {094505}
  (\bibinfo {year} {2012})}\BibitemShut {NoStop}%
\bibitem [{\citenamefont {Kemper}\ \emph {et~al.}(2011)\citenamefont {Kemper},
  \citenamefont {Korshunov}, \citenamefont {Devereaux}, \citenamefont {Fry},
  \citenamefont {Cheng},\ and\ \citenamefont {Hirschfeld}}]{Kemper2011}%
  \BibitemOpen
  \bibfield  {author} {\bibinfo {author} {\bibfnamefont {A.~F.}\ \bibnamefont
  {Kemper}}, \bibinfo {author} {\bibfnamefont {M.~M.}\ \bibnamefont
  {Korshunov}}, \bibinfo {author} {\bibfnamefont {T.~P.}\ \bibnamefont
  {Devereaux}}, \bibinfo {author} {\bibfnamefont {J.~N.}\ \bibnamefont {Fry}},
  \bibinfo {author} {\bibfnamefont {H.-P.}\ \bibnamefont {Cheng}}, \ and\
  \bibinfo {author} {\bibfnamefont {P.~J.}\ \bibnamefont {Hirschfeld}},\ }\href
  {\doibase 10.1103/PhysRevB.83.184516} {\bibfield  {journal} {\bibinfo
  {journal} {Phys. Rev. B}\ }\textbf {\bibinfo {volume} {83}},\ \bibinfo
  {pages} {184516} (\bibinfo {year} {2011})}\BibitemShut {NoStop}%
\bibitem [{\citenamefont {Brouet}\ \emph {et~al.}(2016)\citenamefont {Brouet},
  \citenamefont {LeBoeuf}, \citenamefont {Lin}, \citenamefont {Mansart},
  \citenamefont {Taleb-Ibrahimi}, \citenamefont {Le~F\`evre}, \citenamefont
  {Bertran}, \citenamefont {Forget},\ and\ \citenamefont
  {Colson}}]{Brouet2016}%
  \BibitemOpen
  \bibfield  {author} {\bibinfo {author} {\bibfnamefont {V.}~\bibnamefont
  {Brouet}}, \bibinfo {author} {\bibfnamefont {D.}~\bibnamefont {LeBoeuf}},
  \bibinfo {author} {\bibfnamefont {P.-H.}\ \bibnamefont {Lin}}, \bibinfo
  {author} {\bibfnamefont {J.}~\bibnamefont {Mansart}}, \bibinfo {author}
  {\bibfnamefont {A.}~\bibnamefont {Taleb-Ibrahimi}}, \bibinfo {author}
  {\bibfnamefont {P.}~\bibnamefont {Le~F\`evre}}, \bibinfo {author}
  {\bibfnamefont {F.}~\bibnamefont {Bertran}}, \bibinfo {author} {\bibfnamefont
  {A.}~\bibnamefont {Forget}}, \ and\ \bibinfo {author} {\bibfnamefont
  {D.}~\bibnamefont {Colson}},\ }\href {\doibase 10.1103/PhysRevB.93.085137}
  {\bibfield  {journal} {\bibinfo  {journal} {Phys. Rev. B}\ }\textbf {\bibinfo
  {volume} {93}},\ \bibinfo {pages} {085137} (\bibinfo {year}
  {2016})}\BibitemShut {NoStop}%
\bibitem [{\citenamefont {Yin}\ \emph {et~al.}(2012)\citenamefont {Yin},
  \citenamefont {Haule},\ and\ \citenamefont {Kotliar}}]{Yin2012}%
  \BibitemOpen
  \bibfield  {author} {\bibinfo {author} {\bibfnamefont {Z.~P.}\ \bibnamefont
  {Yin}}, \bibinfo {author} {\bibfnamefont {K.}~\bibnamefont {Haule}}, \ and\
  \bibinfo {author} {\bibfnamefont {G.}~\bibnamefont {Kotliar}},\ }\href
  {\doibase 10.1103/PhysRevB.86.195141} {\bibfield  {journal} {\bibinfo
  {journal} {Phys. Rev. B}\ }\textbf {\bibinfo {volume} {86}},\ \bibinfo
  {pages} {195141} (\bibinfo {year} {2012})}\BibitemShut {NoStop}%
\bibitem [{\citenamefont {Mravlje}\ \emph {et~al.}(2011)\citenamefont
  {Mravlje}, \citenamefont {Aichhorn}, \citenamefont {Miyake}, \citenamefont
  {Haule}, \citenamefont {Kotliar},\ and\ \citenamefont
  {Georges}}]{Mravlje2011}%
  \BibitemOpen
  \bibfield  {author} {\bibinfo {author} {\bibfnamefont {J.}~\bibnamefont
  {Mravlje}}, \bibinfo {author} {\bibfnamefont {M.}~\bibnamefont {Aichhorn}},
  \bibinfo {author} {\bibfnamefont {T.}~\bibnamefont {Miyake}}, \bibinfo
  {author} {\bibfnamefont {K.}~\bibnamefont {Haule}}, \bibinfo {author}
  {\bibfnamefont {G.}~\bibnamefont {Kotliar}}, \ and\ \bibinfo {author}
  {\bibfnamefont {A.}~\bibnamefont {Georges}},\ }\href {\doibase
  10.1103/PhysRevLett.106.096401} {\bibfield  {journal} {\bibinfo  {journal}
  {Phys. Rev. Lett.}\ }\textbf {\bibinfo {volume} {106}},\ \bibinfo {pages}
  {096401} (\bibinfo {year} {2011})}\BibitemShut {NoStop}%
\bibitem [{\citenamefont {Stadler}\ \emph {et~al.}(2015)\citenamefont
  {Stadler}, \citenamefont {Yin}, \citenamefont {von Delft}, \citenamefont
  {Kotliar},\ and\ \citenamefont {Weichselbaum}}]{Stadler2015}%
  \BibitemOpen
  \bibfield  {author} {\bibinfo {author} {\bibfnamefont {K.~M.}\ \bibnamefont
  {Stadler}}, \bibinfo {author} {\bibfnamefont {Z.~P.}\ \bibnamefont {Yin}},
  \bibinfo {author} {\bibfnamefont {J.}~\bibnamefont {von Delft}}, \bibinfo
  {author} {\bibfnamefont {G.}~\bibnamefont {Kotliar}}, \ and\ \bibinfo
  {author} {\bibfnamefont {A.}~\bibnamefont {Weichselbaum}},\ }\href {\doibase
  10.1103/PhysRevLett.115.136401} {\bibfield  {journal} {\bibinfo  {journal}
  {Phys. Rev. Lett.}\ }\textbf {\bibinfo {volume} {115}},\ \bibinfo {pages}
  {136401} (\bibinfo {year} {2015})}\BibitemShut {NoStop}%
\end{thebibliography}%

\end{document}